\documentclass[preprint,12pt]{elsarticle}

 \usepackage{graphicx}

\usepackage{amssymb}

\begin{document}

\begin{frontmatter}



\title{An ultimate storage ring lattice with vertical emittance generated by damping wigglers}


\author{Xiaobiao Huang}
\ead{xiahuang@slac.stanford.edu}
\address{SLAC National Accelerator Laboratory, 2575 Sand Hill Road, Menlo Park, CA 94025}

\begin{abstract}
We discuss the approach of generating round beams for ultimate storage rings using 
vertical damping wigglers (with horizontal magnetic field). The vertical damping wigglers provide 
damping and excite vertical emittance. This eliminates the need to generate large linear coupling 
that is impractical with traditional off-axis injection. 
We use a PEP-X compatible lattice to demonstrate the approach. 
This lattice uses separate quadrupole and sextupole magnets with realistic gradient strengths. 
Intrabeam scattering effects are calculated. 
The horizontal and vertical emittances are 22.3 pm and 10.3 pm, respectively, 
for a 200 mA, 4.5 GeV beam, with a vertical damping wiggler of a total length of 90 meters, 
peak field of 1.5 T and wiggler period of 100 mm. 
\end{abstract}


\end{frontmatter}


\section{Introduction}
In present day third generation light sources, the vertical emittance is usually small compared to 
the horizontal emittance. It is typically a few percent of the latter or below without coupling 
correction and can reach pico-meter level with coupling correction. For ultimate storage rings (USR), it is not 
advisable to maintain the same level vertical-to-horizontal emittance ratio. This is because the 
horizontal emittance will already be diffraction-limited and hence there is no need to make the 
vertical emittance any smaller. In addition, a smaller vertical emittance will cause significant emittance growth 
due to intrabeam 
scattering (IBS) and also severe Touschek beam loss. Most USR designs to-date (such as PEP-X~\cite{PEPXusr}) 
assume the vertical emittance to 
be equal to the horizontal emittance, resulting in a "round beam". 

\begin{figure}[htbp]
   \begin{center}
      \includegraphics[width=0.5\textwidth,]{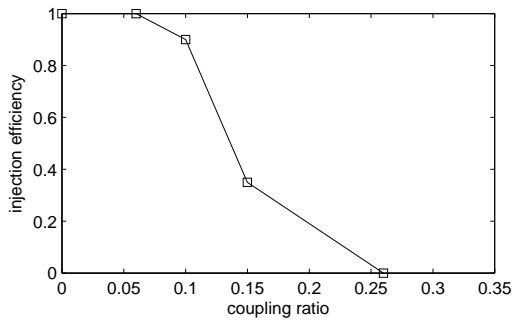} 
      \caption{ Injection efficiency vs. coupling ratio at SPEAR3.  }
      \label{figInjEffCpl}
   \end{center}
\end{figure}

Vertical emittance in a storage ring can be generated with linear coupling or vertical dispersion. 
A round beam can be achieved with 100\% 
linear coupling, in which case the horizontal and vertical emittances are 50\% 
of the natural emittance. The reduction of horizontal emittance by a factor of 2 is 
a significant benefit of this approach. However, large coupling between the two 
transverse directions will cause injection difficulties for off-axis injection. 
The injected beam, initially at a large horizontal offset, will take large vertical 
oscillation and likely get lost to small vertical apertures such as the small-gap 
insertion devices. Effectively, large coupling with small vertical apertures causes the 
dynamic aperture to decrease. This is experimentally demonstrated on the SPEAR3 storage ring 
as is shown in Figure~\ref{figInjEffCpl}, which shows that the injection efficiency drops to zero 
at or before the coupling ratio is increased to 26\%. 
Large linear coupling may also reduce Touschek lifetime since the 
horizontal oscillation of the Touschek particles will be coupled to the vertical plane which 
usually has smaller apertures. 

The second approach to generate vertical emittance is to create vertical dispersion inside 
dipole magnets. This will not cause severe injection and lifetime difficulties, but will 
lose the benefit of horizontal emittance reduction. And since strong skew quadrupoles are needed to create 
large vertical dispersion,  it may be inevitable to introduce large linear coupling. 

We have studied a third approach which can mitigate the negative effects of both of the above 
approaches. In this approach we use vertical damping wigglers (with horizontal magnetic field) to achieve 
both the reduction of horizontal emittance and the generation of vertical emittance. Damping 
wigglers are usually required for USRs because in USRs the dipole bending radius is large and hence the 
radiation energy loss from dipole magnets is too small for sufficient damping, which is required 
for controlling collective effects such as intrabeam scattering and beam instabilities. Usually damping 
wigglers have vertical magnet field that causes wiggling beam motion on the horizontal plane. 
The horizontal dispersion generated by the damping wiggler itself contributes to an increase of 
the horizontal emittance. The relative emittance increase can be significant when the natural emittance is small. 
Choosing to use small period damping wigglers alleviates the emittance growth problem to some extent. 
But it puts a challenge to the damping wiggler design and increases the cost.   
A vertical damping wiggler does not increase the horizontal emittance and in the same 
time generates the desirable vertical emittance. Therefore it is reasonable to use vertical 
damping wigglers for USRs. 
The idea of using vertical damping wigglers to generate vertical emittance has been independently proposed in 
Refs.~\cite{APnote51,Bogomyagkov}. 

In this study we demonstrate this approach with a lattice that is compatible with the PEP 
tunnel at SLAC National Accelerator Laboratory. 
In section~\ref{secEstimate} we use a simple model to calculate and compare the 
emittances with the horizontal and vertical damping wiggler approaches. 
In section~\ref{secPEPXlat} the PEP compatible lattice with vertical damping wigglers is presented. 
Emittance parameters with intrabeam scattering effects are given in section ~\ref{secIBS}. 
The conclusions are given in section~\ref{secConclusion}. 

\section{Theoretic calculation}\label{secEstimate}
The effects of vertical damping wigglers can be analytically estimated. Suppose the 
wiggler peak field is $B_w$,  its length is $L_w$, and the horizontal field is given by
\begin{eqnarray}\label{eqBfieldDW}
B_x &=& B_w \cosh kx \cos ks, \qquad B_y = 0,
\end{eqnarray}
where $k=2\pi/\lambda_w$ and $\lambda_w$ is the wiggler period, then the vertical closed orbit inside the damping wiggler (DW) is
~\cite{SYLeeBook}
\begin{eqnarray}
y_{co} &=& \frac1{\rho_w k^2}(1-\cos ks), \qquad y'_{co} =\frac1{\rho_w k}\sin ks,
\end{eqnarray}
where $\rho_w=B\rho/B_w$ is the minimum bending radius. The vertical dispersion generated by 
the DW itself is 
\begin{eqnarray}
D_y &=& -\frac1{\rho_w k^2}(1-\cos ks), \qquad D'_y =-\frac1{\rho_w k}\sin ks,
\end{eqnarray}
Consequently the radiation integral contributions are 
\begin{eqnarray}\label{eqIntDW}
I_{2w} &=& \frac{L_w}{2\rho^2_w}, \qquad 
I_{3w} = \frac{4 L_w}{3 \pi \rho^3_w}, \nonumber \\ 
I_{4wy} &=& \frac{3 L_w}{8 \pi \rho^4_w k^2}, \quad 
I_{5wy} = \frac{4 <\beta_y> L_w}{15 \pi \rho^5_w k^2},  
\end{eqnarray}
where $<\beta_y>$ is the average vertical beta function across the DW. 
The emittances and momentum spread are given by 
\begin{eqnarray}\label{eqEmmitxy}
\sigma^2_\delta &=& \gamma^2 C_q \frac{I_3+I_{3w}}{I_2+I_{2w}}\frac1{2+\mathcal{D}_x+\mathcal{D}_y}, \\
\epsilon_x &=& \gamma^2 C_q \frac{I_5}{I_2+I_{2w}}\frac1{1-\mathcal{D}_x}, \\
\epsilon_y &=& \gamma^2 C_q \frac{I_{5wy}}{I_2+I_{2w}}\frac1{1-\mathcal{D}_y}, 
\end{eqnarray}
where $I_{2-5}$ are radiation integrals for the bare lattice and 
\begin{eqnarray}
\mathcal{D}_x &=& \frac{I_4}{I_2+I_{2w}}, \qquad 
\mathcal{D}_y = \frac{I_{4wy}}{I_2+I_{2w}}.
\end{eqnarray}

\begin{figure}[htbp]
   \begin{center}
      \includegraphics[width=0.40\textwidth]{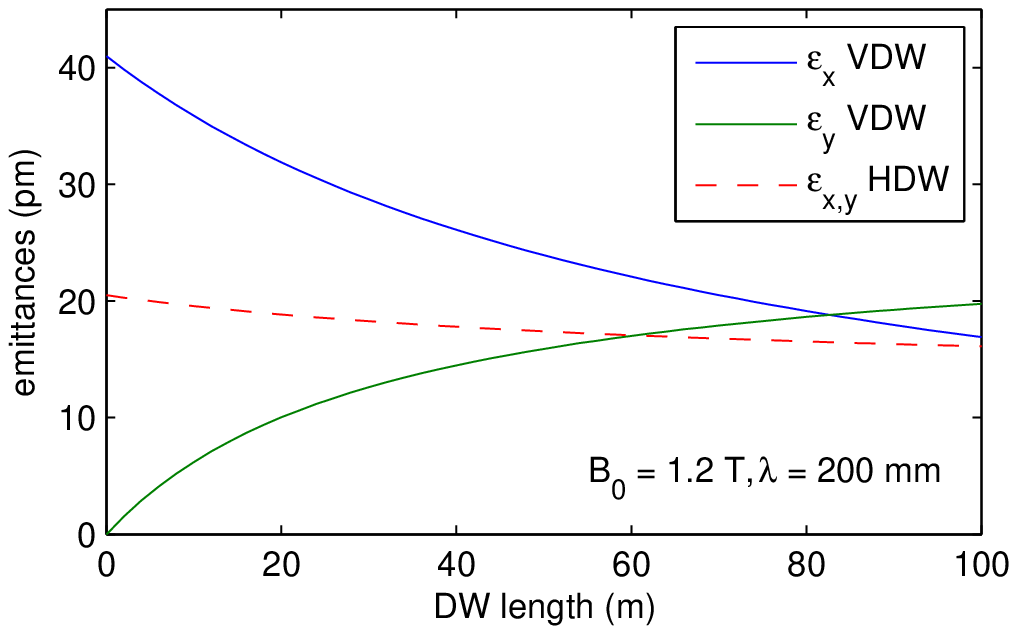} 
      \includegraphics[width=0.40\textwidth]{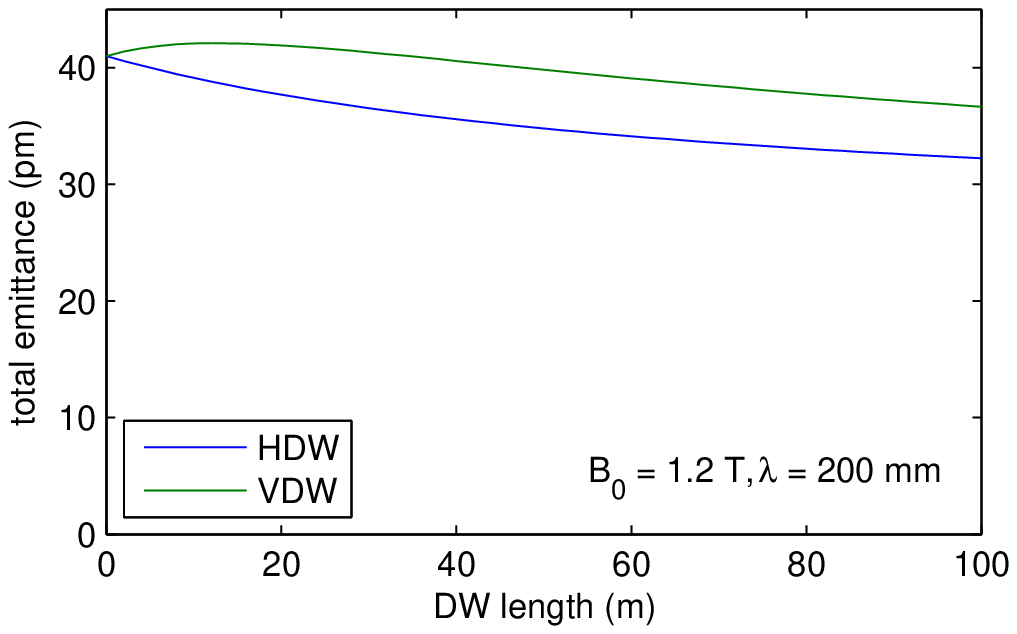} 
      \includegraphics[width=0.40\textwidth]{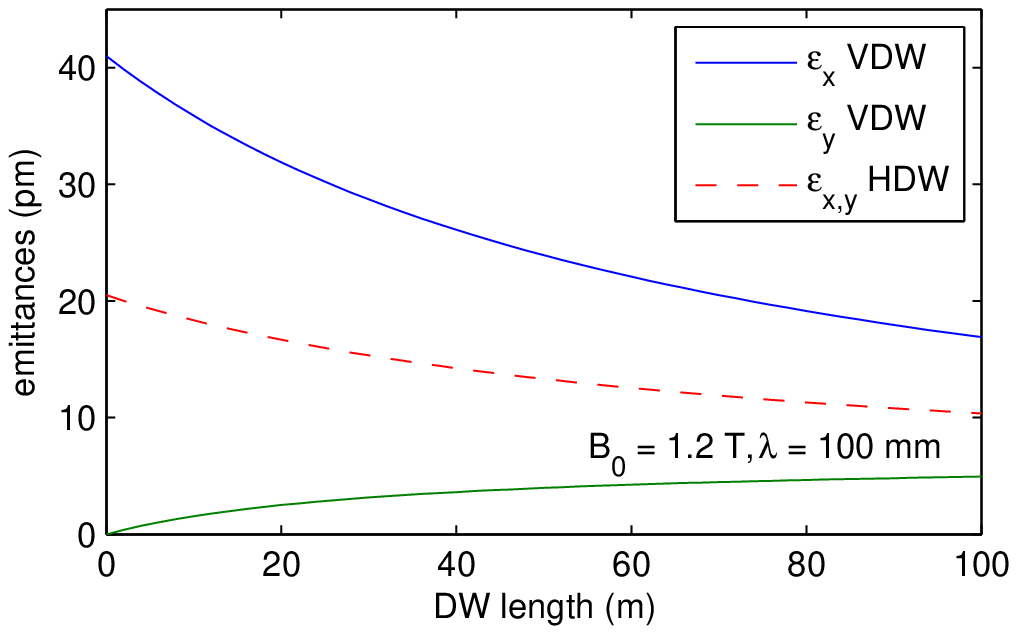} 
      \includegraphics[width=0.40\textwidth]{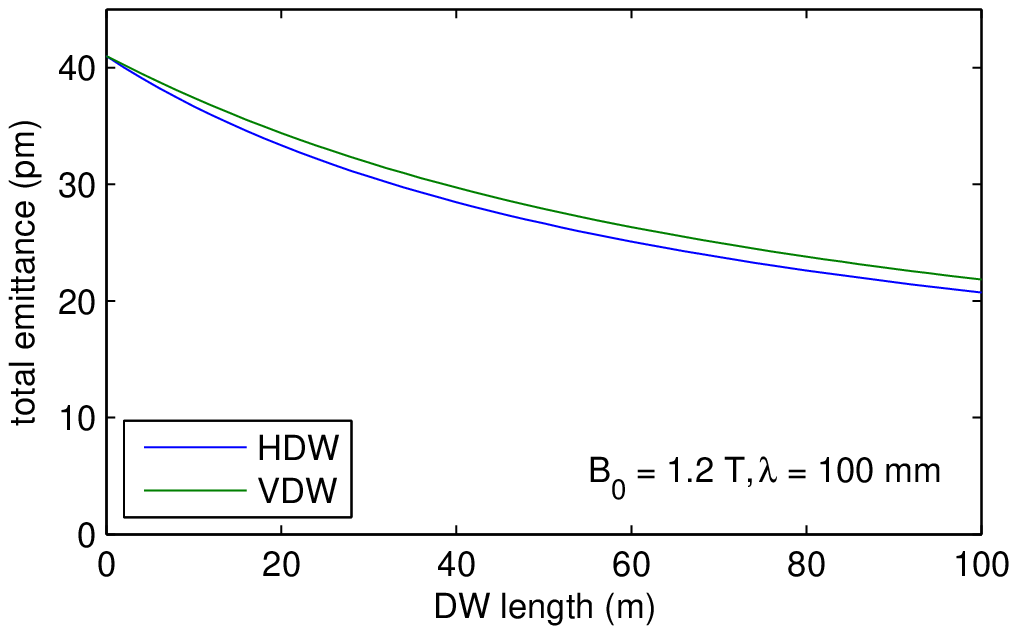} 
      \includegraphics[width=0.40\textwidth]{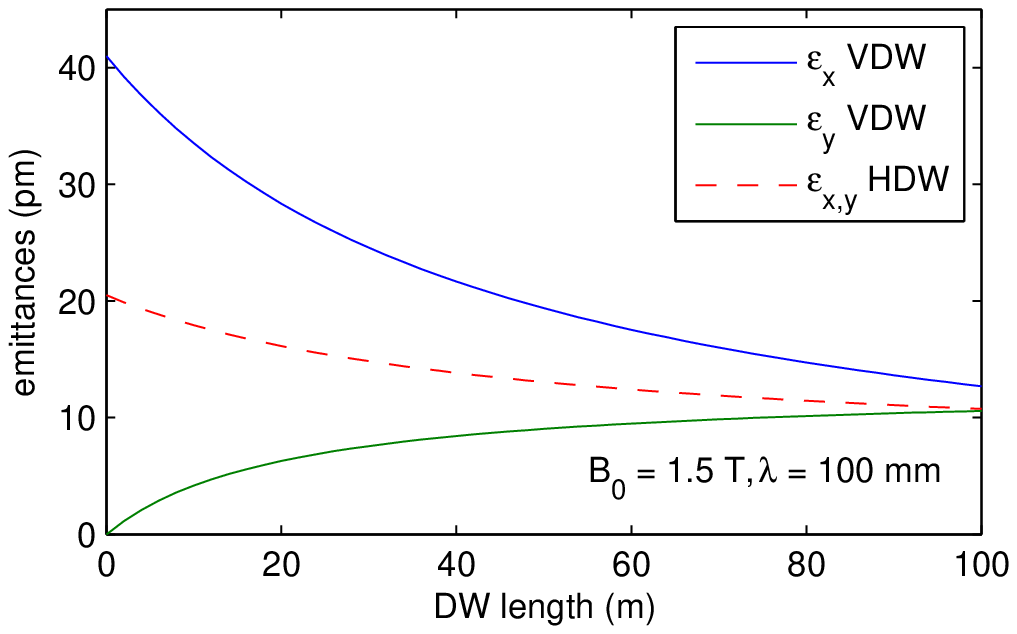} 
      \includegraphics[width=0.40\textwidth]{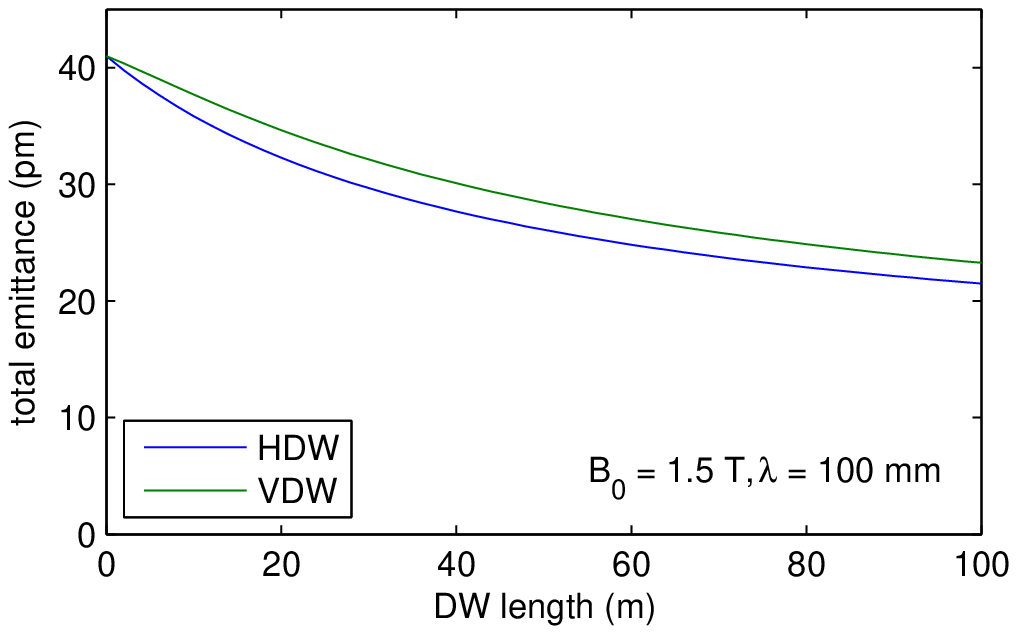} 
      \caption{ Comparison of the emittances of a PEP-X ring with vertical or horizontal damping wigglers. 
      Top row with $B_w=1.2$ T and $\lambda_w=200$~mm; 
      middle row with $B_w=1.2$ T and $\lambda_w=100$~mm;
      bottom tow with $B_w=1.5$ T and $\lambda_w=100$~mm. 
      A 100\% 
      coupling is assumed for the regular horizontal DW case.   }
      \label{figEmittxyCmp1}
   \end{center}
\end{figure}

We now consider a PEP-X compatible lattice at 4.5 GeV (see section~\ref{secPEPXlat}). The relevant radiation integrals without DWs are 
\begin{eqnarray}
I_2 &=& 0.1026 \,{\rm m}^{-1}, \quad  \,   I_3 =  1.674\times10^{-3}  \,{\rm m}^{-2}, \nonumber \\ 
I_4 &=&  -0.1215 \,{\rm m}^{-1}, \quad  I_5 = 3.092\times10^{-7} \,{\rm m}^{-1}. 
\end{eqnarray}
Assuming the average beta function 
over the DW is 10 m, 
the emittances as a function of wiggler length is calculated and compared to 
the case with a regular horizontal damping wiggler 
for various sets of peak magnetic field and wiggler period values. 
The results are shown in Figure \ref{figEmittxyCmp1}. 
Clearly the vertical DW provides damping of the horizontal emittance and in the meantime generates 
vertical emittance. The total emittance is only slightly larger than the case with a regular horizontal 
DW. The difference is smaller for smaller wiggler periods. 

\section{Application to a PEP-X compatible lattice}\label{secPEPXlat}
\begin{figure}[htbp]
   \begin{center}
      \includegraphics[width=0.4\textwidth,,angle=-90]{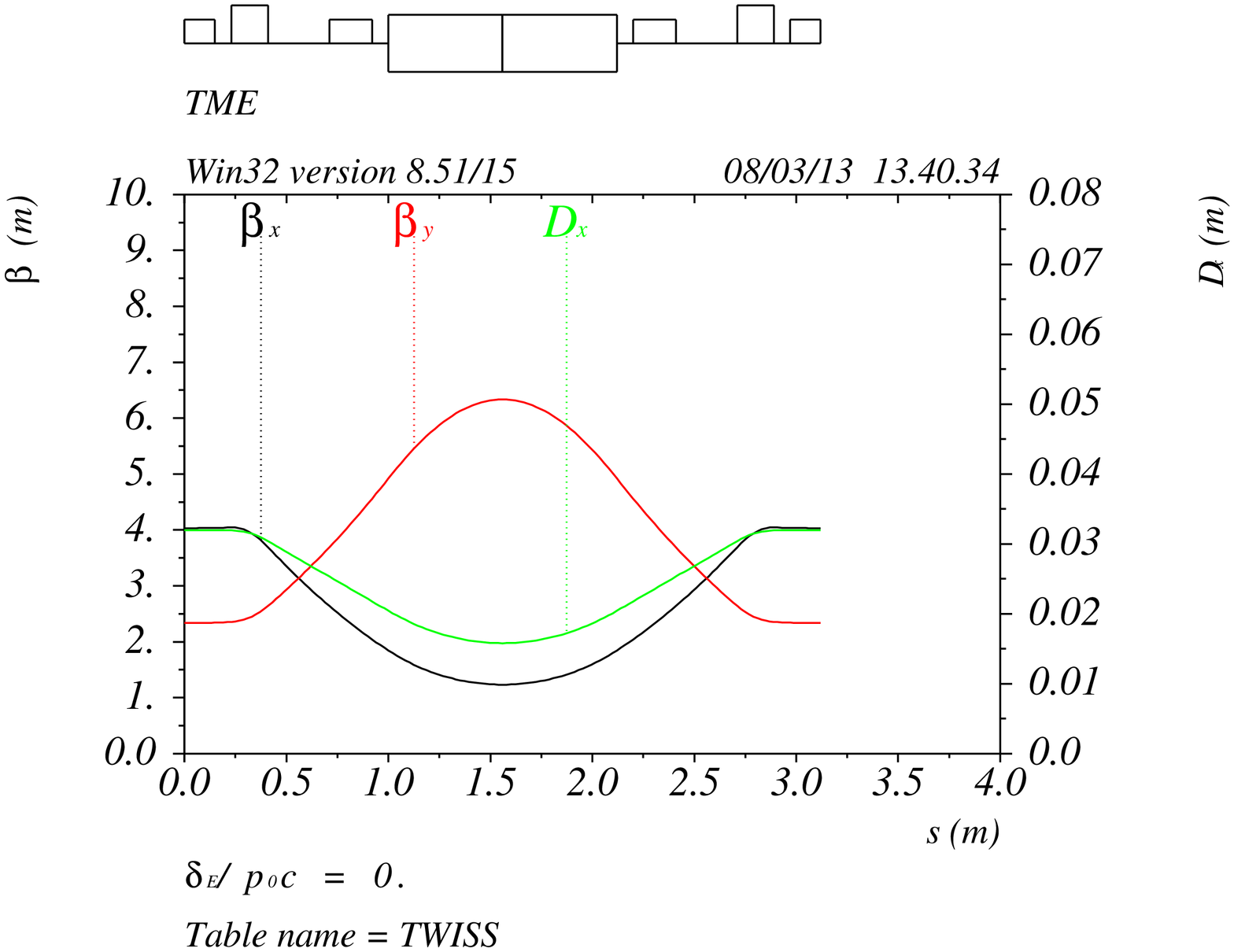}  
      \includegraphics[width=0.4\textwidth,,angle=-90]{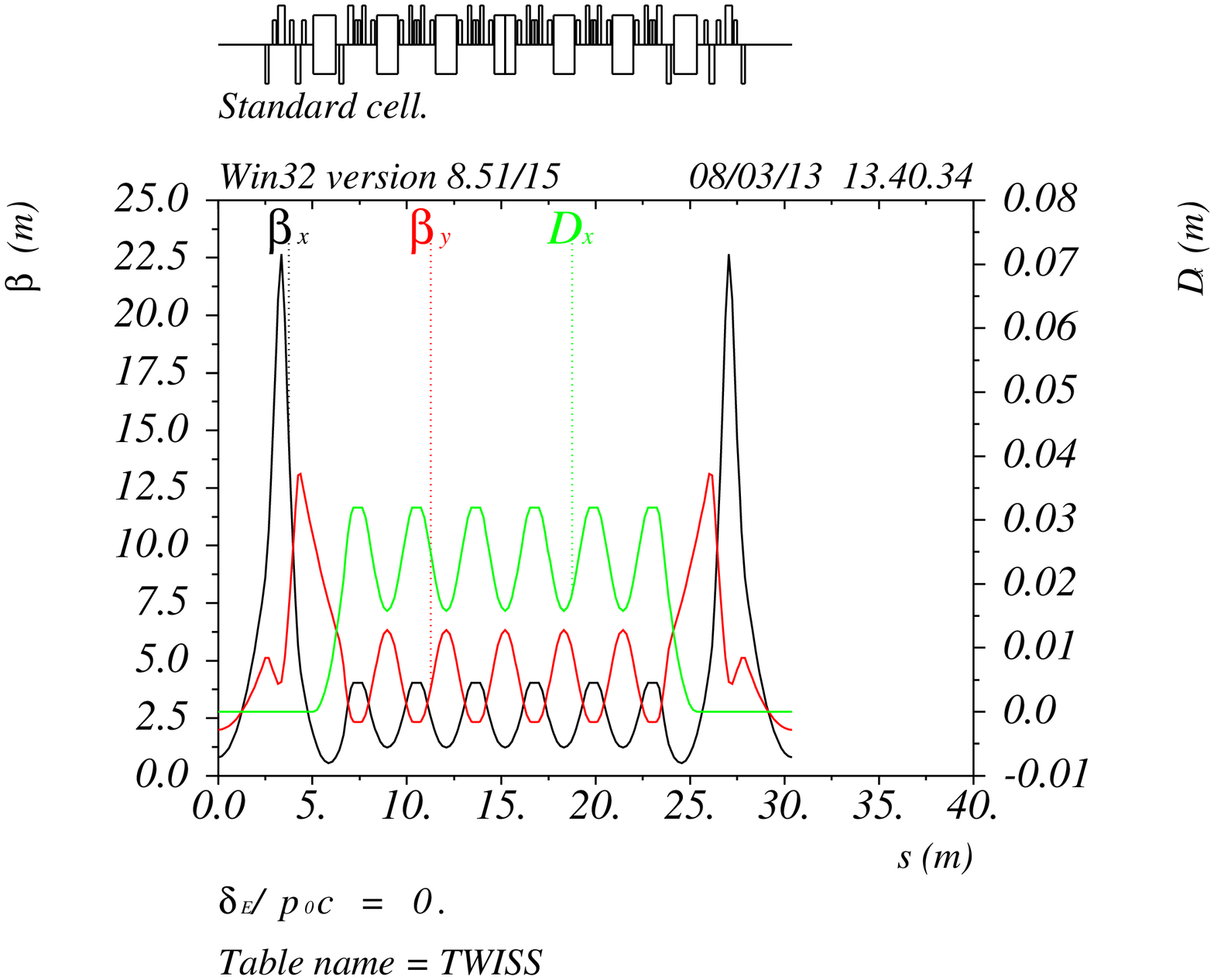} 
      \caption{ The TME cell (left) and 7BA cell (right) for the PEP-X compatible lattice.   }
      \label{figTMECell}
   \end{center}
\end{figure}

We have implemented the vertical DW approach for a PEP-X compatible lattice with the design 
beam energy at 4.5~GeV. 
This lattice is similar to the PEP-X USR design as it adopts the 
same MBA and fourth order achromat approach~\cite{PEPXusr}. 
The 2.2-km long PEP tunnel has a hexagonal geometry. There are six 120-m long straight sections 
which can be used to host long damping wigglers. 
The lattice has 6 arcs, each 
consists of 8 MBA (with $M=7$) cells. An MBA cell is composed of 5 identical TME cells in the middle and 
two matching cells at the ends. The MBA cell and the TME cell  
are shown in Figure~\ref{figTMECell}. The TME and 7BA cell lengths are 
3.12~m and 30.4~m, respectively. 
The TME dipole magnet is 1.12~m in length and its bending angle is 1.0475~$^{\circ}$. 
This dipole is a combined-function magnet with a defocusing 
quadrupole component and the normalized gradient is $-0.7989$~m$^{-2}$. 
The focusing quadrupole (QF) is split into two halves to put the SF sextupole 
in between. The length of each half is 0.18~m. The length of SF is 0.30~m. 
One SD sextupole magnet is put at each end of the dipole. Its length is 0.21~m. 
The matching dipole has no quadrupole gradient. Its length is 8\%
longer than the TME dipole. 
At each end of the MBA cell, outside of the matching dipole, there is a 
quadrupole triplet. Three harmonic sextupoles are put between these magnets.
The minimum edge-to-edge distance for magnets is 8~cm to accommodate coils and BPMs~\cite{MAXIV}. 
The quadrupole strength is below 51~T/m and the sextupole 
strength is below 7500~T/m$^{2}$. With a bore radius of 12.5~mm, the pole tip 
magnetic field would be below 0.64~T for quadrupoles and below 0.59~T for sextupoles. 

The insertion device straight sections between the MBA cells are 5 meter long. The horizontal and vertical 
beta functions at the centers of these straight sections are 0.8~m and 2.0~m, respectively. 
The horizontal beta function is made very small to provide better matching of the electron and photon optics. 
But we keep the vertical beta at a level close to half of the ID straight length to 
allow small gap insertion devices~\cite{Rabedeau}. 

\begin{figure}[htbp]
   \begin{center}
      \includegraphics[width=0.4\textwidth,angle=-90]{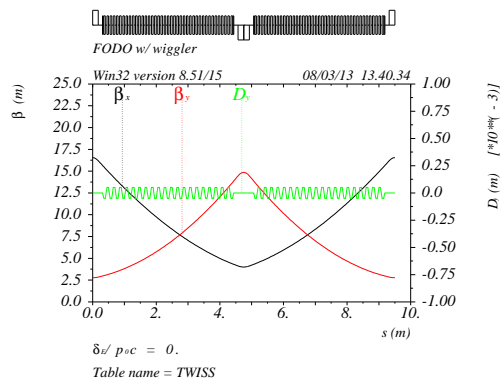}
      \caption{ One FODO lattice period with damping wigglers. }
      \label{figBetaDWOne}
   \end{center}
\end{figure}

\begin{figure}[htbp]
   \begin{center}
      \includegraphics[width=0.4\textwidth,,angle=-90]{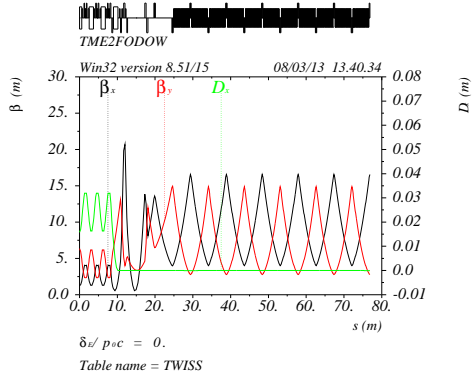}
      \caption{ Half of the DW straight section.  }
      \label{figBetaDW}
   \end{center}
\end{figure}

The 120-m long straight sections are filled with FODO cells. 
One of the long straight sections houses the damping wigglers. The  
wiggler sections are 4.06 meter long and are put between the quadrupoles of the FODO cells. 
The optics functions are shown in Figure \ref{figBetaDWOne} for a FODO cell for the case 
with wiggler period at 200~mm and peak field at 1.2~T. 
Optics function for one half of the long straight section is as shown in Figure \ref{figBetaDW}.

The ring lattice parameters for three wiggler settings are compared in Table \ref{tabRingPara}. 
The parameters were calculated with MAD8~\cite{MAD8}. 
The results agree with the prediction given in Figure \ref{figEmittxyCmp1}. 
For the vertical DW sets of (1.2~T, 200~mm) and (1.5~T, 100~mm), the horizontal and 
vertical emittances are nearly equal, with values down to 17/17 pm and 13/10 pm, respectively. 
\begin{table}[!hbt] 
\caption{Ring parameters with or without vertical damping wigglers }
\label{tabRingPara}
 \begin{center}  
  \begin{tabular*}{0.8\textwidth}%
     {@{\extracolsep{\fill}}l|cccc}
  \hline
  parameters & no DW & VDW-1 & VDW-2 & VDW-3\\
  \hline  
  Energy (GeV) & \multicolumn{4}{c}{4.5}  \\
  Circumference (m) & \multicolumn{4}{c}{2199.3}  \\
  $\nu_{x,y}$ & \multicolumn{4}{c}{130.15/73.30} \\
  $\alpha_c$  & \multicolumn{4}{c}{0.38$\times 10^{-4}$} \\
  \hline
  VDW length (m) & 0   &   90.8  &  89.4 & 89.4\\
  VDW $B_w$ (T) &      &   1.2    & 1.2 & 1.5\\
  VDW $\lambda_w$ (m) &      &   0.2    & 0.1 & 0.1 \\
  \hline
  $U_0$ (MeV) &   0.59  & 2.30  & 2.26 & 3.19\\
  $\epsilon_x$ (pm) &   36.5  &  16.5  & 16.7 & 12.7\\
  $\epsilon_y$ (pm) &   0   &  17.1  & 4.3 & 10.2\\
  $\sigma_\delta$ ($\times 0.001$) &  0.77 & 1.05  & 1.05 & 1.10 \\
  damping time $\tau_x$ (ms) & 47 &  21  &  22 & 17\\
  damping time $\tau_y$ (ms) & 112 &  29  & 29 & 21 \\
  damping time $\tau_s$ (ms) & 176 &  17  & 18 &12 \\
  \hline
  \end{tabular*}
  \end{center}
\end{table}  

\section{Intrabeam scattering calculation}\label{secIBS}
Intrabeam scattering (IBS) can significantly increase the emittance and energy spread for 
very low emittance beams at high current. 
To examine how the IBS effects may differ for the two approaches of generating vertical emittance, 
i.e., with full coupling or with vertical damping wiggler, we did IBS calculation for both cases for 
the PEP-X compatible lattice with the high energy approximation model~\cite{PEPXusr,BaneIBS}. 
Similar to PEP-X, we assume the Coulomb log function is $({\mathrm log})=11$. 
For the full coupling case, a horizontal damping wiggler is put into the model to reduce emittance. 
The wiggler parameters are the same as the vertical damping wiggler. For the case corresponding to 
VDW-3 in Table~\ref{tabRingPara} (i.e.,
with peak field 1.5~T, wiggler period 100~mm and wiggler length 89.4~m), 
the emittances are 10.5~pm for both planes.   
For the vertical wiggler case, a small linear coupling ratio of 0.001 is assumed. 
The vertical emittance is almost entirely generated with the vertical damping wiggler. 

\begin{figure}[!htbp]
   \begin{center}
      \includegraphics[width=0.40\textwidth]{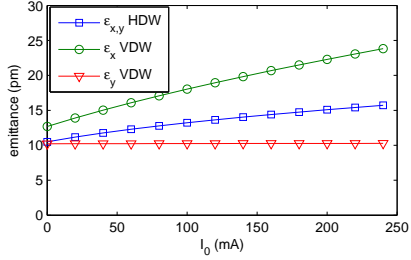} 
      \caption{ Emittance growth vs total beam current, assuming a uniform current 
      distribution in 3300 bunches and a bunch length of 2.7~mm.   }
      \label{figEmitxyvsItotIBS}
   \end{center}
\end{figure}

The emittances vs. beam current for the two cases with IBS effects are shown in Figure~\ref{figEmitxyvsItotIBS}.
The bunch length is assumed to be $\sigma_z=2.7$~mm, corresponding to 
an RF gap voltage of 6~MV with the 476.0 MHz rf system. 
The total number of bunches is assumed to be 3300. 
For the vertical DW case, the horizontal emittance has a significant increase. But the vertical emittance growth is 
very small. This is because the vertical dispersion is confined to inside the damping wiggler, which 
constitutes only a small fraction of the ring, while the horizontal dispersion is present at all arc 
areas. In addition, the vertical dispersion is much smaller than the horizontal dispersion 
while the horizontal and vertical emittances at zero current are nearly the same. This is because the average 
bending field in the damping wiggler is much stronger than in the bending magnets. 
Overall the vertical IBS growth rate is much smaller than the horizontal plane because 
the IBS growth rate is proportional to the dispersion invariant averaged over the ring circumference. 

\begin{figure}[!htbp]
   \begin{center}
      \includegraphics[width=4in]{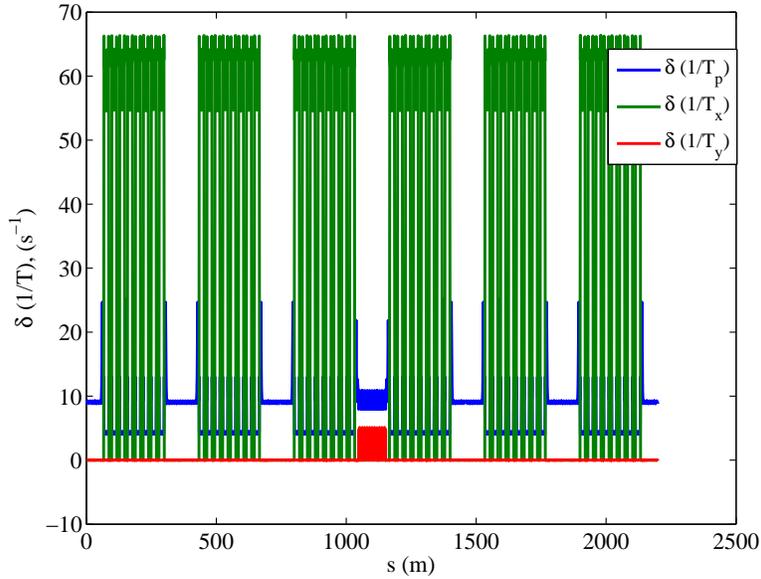} 
      \caption{ The distribution of local IBS growth rate for the three dimensions for a beam current of 200~mA 
      in 3300 bunches.  }
      \label{figIBSdvT}
   \end{center}
\end{figure}

The distribution of the IBS growth rate for the vertical DW case 
is shown in Figure~\ref{figIBSdvT} for the case with a 200~mA total current. 
For this case, the average IBS growth rates for $x$, $y$, $p$ directions are 26.1~s$^{-1}$,
0.20~s$^{-1}$ and 8.4~s$^{-1}$, respectively. 
The corresponding emittances are $\epsilon_x = 22.3$~pm and $\epsilon_y = 10.3$~pm and the momentum spread
is $\sigma_\delta = 1.16\times 10^{-3}$. 
If harmonic cavities are used to lengthen the bunch to $\sigma_z=5$~mm, the $x$, $y$, $p$ IBS growth rates 
become 18.8~s$^{-1}$, 0.13~s$^{-1}$ and 5.5~s$^{-1}$, respectively, and the emittances and the 
momentum spread become $\epsilon_x = 18.4$~pm, $\epsilon_y = 10.25$~pm and $\sigma_\delta = 1.13\times 10^{-3}$. 

\section{Conclusion}\label{secConclusion}
For ultimate storage rings that require off-axis injection, we propose the use of 
vertical damping wiggler (with horizontal magnetic field) to generate vertical emittance 
in order to obtain round beams. 
This approach is better than generating round beams with 100\% 
coupling because it does not couple the large amplitude horizontal oscillation of the injected beam 
to the vertical plane and therefore the small vertical apertures in the ring does not pose severe limitation to the 
dynamic aperture. 
It is shown that for damping wigglers with reasonably small wiggler period (e.g., $\lambda_w = 100$ mm), 
the total emittances of the two approaches are nearly equal. 

A PEP-X compatible lattice is designed to demonstrate this approach. It consists of 6 arcs, each made up of 
8 MBA cells. The beam energy is assumed to be 4.5 GeV. 
The quadrupoles and sextupoles are separate function magnets, with 
strengths below 51 T/m and 7500 T/m$^2$, respectively. 
The bare lattice horizontal emittance is 36.5 pm.
The beta functions at the middle of straights are 0.8 m horizontal and 2.0 m vertical, which allows  
good matching to the photon optics and supports the use of small 
gap insertion devices. 
  
When a 90-m long vertical damping wiggler with peak field at 1.5 T and wiggler period at 
100 mm is put into one of the long straight sections, 
the horizontal and vertical emittances are 13~pm and 10~pm, respectively. 
The rms momentum spread is 1.1$\times 10^{-3}$. 
Intrabeam scattering is calculated for this case. For 200 mA beam current in 3300 bunches and a bunch length of $\sigma_z=2.7$~mm, 
the horizontal and vertical emittance become 22.3 pm and 10.3 pm, respectively. 
The rms momentum spread is $1.16\times 10^{-3}$.

\section*{Acknowledgment}
The study is supported by DOE Contract No. DE-AC02-76SF00515.









\end{document}